\documentclass{article}
\usepackage{spconf,graphicx,cite}

\usepackage{array,amssymb,amsmath,amsfonts}
\usepackage{subfigure}
\usepackage{multirow}
\usepackage{math}
\usepackage{tikz}
\usetikzlibrary{arrows,automata,shapes}
\usepackage{bm,color}% bold math

\newtheorem{theorem}{Theorem}

\title{Rectified Binaural Ratio: A Complex T-Distributed\\ Feature for Robust Sound Localization}
\name{Antoine Deleforge${}^*$ and Florence Forbes${}^\dagger$ }
\address{${}^*$Inria Rennes - Bretagne Atlantique\hspace{5mm}
         ${}^\dagger$Inria Grenoble - Rh\^one-Alpes \; \;(firstname.lastname@inria.fr)}

\begin{document}

\maketitle
\vspace{-20mm}
\begin{abstract}
\vspace{-2mm}
Most existing methods in binaural sound source localization rely on some kind of aggregation of phase- and level- difference cues in the time-frequency plane. While different aggregation schemes exist, they are often heuristic and suffer in adverse noise conditions. In this paper, we introduce the \textit{rectified binaural ratio} as a new feature for sound source localization. We show that for Gaussian-process point source signals corrupted by stationary Gaussian noise, this ratio follows a complex t-distribution with explicit parameters. This new formulation provides a principled and statistically sound way to aggregate binaural features in the presence of noise. We subsequently derive two simple and efficient methods for robust relative transfer function and time-delay estimation. Experiments on heavily corrupted simulated and speech signals demonstrate the robustness of the proposed scheme.
\end{abstract}
\vspace{-4mm}
\section{Introduction}
\vspace{-4mm}
The most widely used features for binaural (two microphones) sound source localization are the measured time delays and level differences between the two microphones. For a single source signal in the absence of noise, these features correspond in the frequency domain to the ratio of the Fourier transforms of the right- and the left-microphone signals. This ratio is called the \textit{relative transfer function} (RTF) \cite{gannot2001signal}, and only depends on the source's spatial characteristics, \textit{e.g.}, its position relative to the microphones.
The log-amplitudes and phases of the RTF are referred to as \textit{interaural level differences} (ILD) and \textit{interaural phase differences} (IPD) in the binaural literature. Many binaural sound source localization methods rely on some kind of aggregation of these cues over the time-frequency plane \cite{knapp1976generalized,Aarabi02,mandel2010model,woodruff2012binaural,zohny2014modelling,mandel2015enforcing,deleforge2015acoustic}. The generalized cross-correlation (GCC) method \cite{knapp1976generalized} consists of weighting the cross-power spectral density (CPSD) of two signals in order to estimate their delay in the time-domain (CPSD phases and IPD are the same). A successful GCC method is the phase transform (PHAT), in which IPD cues are equally weighted. The popular sound localization method PHAT-histogram aggregates these cues using histograms \cite{Aarabi02}. In \cite{woodruff2012binaural}, a heuristic binaural cue weighting scheme based on signals' onsets is proposed. In \cite{mandel2010model}, both ILD and IPD cues are modeled as real Gaussians and their frequency-dependent variances are estimated through an expectation-maximization (EM) procedure referred to as MESSL. A number of extensions of MESSL have later been developed \cite{zohny2014modelling,mandel2015enforcing,deleforge2015acoustic}, including one using t-distributions for ILD and IPD cues instead of Gaussian distributions \cite{zohny2014modelling}.

While all these methods rely on a weighting scheme of binaural cues, none of these schemes is based on the statistical properties of the source and noise signals. Though, intuitively, a low signal-to-noise-ratio (SNR) at microphones means that a specific cue is less reliable, while a high SNR means that this cue should be given more weight.
In this paper, we prove that the ratio of two complex circular-symmetric Gaussian variables follows a complex t-distribution with explicit parameter expressions. In particular, for the binaural recording of a Gaussian-process source corrupted by stationary Gaussian noise, we show that the mean of the microphone signals' ratio does not only depend on the clean ratio but also on the source and noise statistics. This observation naturally leads to the definition of a new binaural feature referred to as the \textit{rectified binaural ratio} (RBR). The explicit distribution of RBR features provides a principled and statistically sound way of weighting and aggregating them. Based on this, we derive two simple and efficient methods for relative transfer function and time-delay estimation, and test their robustness on heavily corrupted binaural signals.

\vspace{-4mm}
\section{A complex-t model for binaural cues}
\vspace{-4mm}
\label{sec:model}
In the complex short-time Fourier domain, we consider the following model for a binaural setup recording a static point sound source in the presence of noise:
\begin{align}
&\left\{
 \begin{array}{l}
 m_1(f,t)=h_{\textrm{1}}(f,\thetavect)s(f,t) + n_{\textrm{1}}(f,t) \\
 m_2(f,t)=h_{\textrm{2}}(f,\thetavect)s(f,t) + n_{\textrm{2}}(f,t) \\
 \end{array}
\right., \nonumber \\
\label{eq:audio_model}
%&\Leftrightarrow
& \mbox{or equivalently } \; 
 \mvect(f,t) = \hvect(f,\thetavect)s(f,t) + \nvect(f,t).
\end{align}
Here, $(f,t)$ is the frequency-time indexing, $\thetavect$ is a vector of source spatial parameters, \textit{e.g.}, the source position, $\mvect(f,t)=[m_1(f,t) , m_2(f,t)]^\top\in\mathbb{C}^2$ denotes the microphone signals, $s(f,t)\in\mathbb{C}$ denotes the source signal of interest, $\nvect(f,t)=[n_1(f,t) , n_2(f,t)]^\top\in\mathbb{C}^2$ denotes the noise signals and $\hvect(f,\thetavect)=[h_1(f,\thetavect) , h_2(f,\thetavect)]^\top\in\mathbb{C}^2$ denotes the acoustic transfer function from the source to the microphones.
The function  $\hvect(f,\thetavect)$ is of particular interest because it depends on the source position $\thetavect$ but does not depend on the time-varying source and noise signals. Under noise-free and non-vanishing source assumptions, \textit{i.e.} $\nvect(f,t)=\zerovect$ and $s(f,t)\ne0$, it is easily seen that the \textit{binaural ratio} $m_2(f,t)/m_1(f,t)$ is equal to $h_2(f,\thetavect)/h_1(f,\thetavect)=r(f,\thetavect)$, which only depends on the source position. This ratio can hence be used for sound source localization. The quantity $r(f,\thetavect)$ is called \textit{relative transfer function} (RTF) \cite{gannot2001signal}. Its log-amplitudes and phases are respectively referred to as interaural level and phase differences (ILD and IPD).

In practical situations including noise, the ratio $m_2(f,t)/$ $m_1(f,t)$ does no longer depend on $\thetavect$ only, but also on the source and noise signals $s(f,t)$ and $\nvect(f,t)$. These signals are assumed independent, and we consider the following probabilistic models:
\begin{align}
\label{eq:stat_model_s}
P(s(f,t))      &= \mathcal{CN}_1(s(f,t);0,\sigma_s^2(f,t)), \\
\label{eq:stat_model_n}
P(\nvect(f,t)) &= \mathcal{CN}_2(\nvect(f,t);\zerovect,\Rmat_{nn}(f)),
\end{align}
where $\mathcal{CN}_p$ denotes the $p$-variate complex circular-symmetric normal distribution, or \textit{complex-normal}. Its density is \cite{fuhrmann1997complex}:
\begin{equation}
 \mathcal{CN}_p(\xvect;\cvect,\Sigmamat) = \frac{1}{\pi^p|\Sigmamat|}\exp\left(-(\xvect-\cvect)^{\textrm{H}}\Sigmamat^{-1}(\xvect-\cvect)\right), \nonumber
\end{equation}
where $\{\cdot\}^{\textrm{H}}$ denotes the Hermitian transpose. We assume that $\Rmat_{nn}(f)$ is known and constant over time, \textit{i.e.}, noise signals are stationary. However, they are not necessarily pairwise independent and may thus include other point sources. On the other hand, the source signal is a Gaussian process with time-varying variance $\sigma_s^2(f,t)$. This general model is widely used in audio signal processing, in particular for sound source separation, \textit{e.g.}, \cite{vincent2009underdetermined}.
We now introduce the univariate \textit{complex t-distribution} denoted $\mathcal{CT}_1$:
\begin{equation}
\label{eq:complex-t}
 \mathcal{CT}_1(y;\mu,\lambda^2,\nu) = \frac{1}{\pi\lambda^2}\left(1+\displaystyle\frac{|y-\mu|^2}{\nu\lambda^2}\right)^{-(1+\nu)},
\end{equation}
where $\mu \in \mathbb{C}$, $\lambda^2 \in \mathbb{R}^+ $ and $\nu \in \mathbb{R}^+$ are respectively referred to as the mean, spread and degrees of freedom parameters. This definition follows 
a construction of multivariate extensions for the t-distribution \cite{KotzNadarajah2004} applied to the complex plane.
% The density in (\ref{eq:complex-t}) can also be seen as the following real bivariate t-distribution $\mathcal{T}_2$ on the real and imaginary parts:
% \begin{equation}
% \label{eq:complex-t2}
%  \mathcal{CT}_1(y;\mu,\lambda^2,\nu) = \mathcal{T}_2(y ;\mu, \frac{1}{2} \lambda^2 \Imat_2 , 2\nu)
% \end{equation}
% where $y$ and $\mu$ are seen as bivariate vectors and $\Imat_2$ is the $2\times 2$ identity matrix. We can then easily see that assuming (\ref{eq:complex-t}),   the real and imaginary parts are  assumed non correlated and of equal spread parameter $\lambda^2/2$. 
In the real case, the t-distribution arises from the ratio of a Gaussian over the square root of a Chi-square distribution. In the complex case,
%By reformulating the main result of \cite{baxley2010complex} in terms of this new complex-t distribution, we have the following theorem:
we alternatively show the following result:
\begin{theorem}
%[from \cite{baxley2010complex}}
\label{TH:MAIN}
Let $\mvect=[m_1 , m_2]^\top$ be a vector in $\mathbb{C}^2$ following a complex-normal distribution such that
$$
P(\mvect) = \mathcal{CN}_2\biggl(\mvect;\zerovect,
\left[
 \begin{array}{cc}
   \sigma_{m_1}^2 & \rho\sigma_{m_1}\sigma_{m_2} \\
    \rho^*\sigma_{m_1}\sigma_{m_2} & \sigma_{m_2}^2  \\
 \end{array}
\right]
\biggr).
$$
Then the ratio variable $y=m_2/m_1$ follows a complex-t distribution such that
\begin{equation}
P(y) = \mathcal{CT}_1\left(y;\frac{\sigma_{m_2}}{\sigma_{m_1}}\rho^*,\frac{\sigma_{m_2}^2}{\sigma_{m_1}^2}(1-|\rho|^2),1\right).
\end{equation}
\end{theorem}
Here, $\rho=\mathbb{E}\{m_1m_2^*\}/(\sigma_{m_1}\sigma_{m_2})$ is the correlation coefficient between $m_1$ and $m_2$ and $(.)^*$ denotes the complex conjugate. 
This result is consistent with that in \cite{baxley2010complex} but we provide a simpler proof with better insight in Appendix \ref{app:theo1}. 
Theorem \ref{TH:MAIN} can be directly applied to obtain an explicit distribution for the binaural ratio $m_2(f,t)/m_1(f,t)$ under the model defined by (\ref{eq:audio_model}), (\ref{eq:stat_model_s}) and (\ref{eq:stat_model_n}). However, both the mean and the spread of this distribution depend on the noise correlation and variances as well as the transfer functions in a way which is difficult to handle.
We will therefore design a more convenient and somewhat more natural binaural feature by first \textit{whitening} the noise signals in each observed vectors $\mvect(f,t)$, \textit{i.e.}, making them independent and of unit variance. Since $\Rmat_{nn}(f)$ is positive semi-definite, it has a unique positive semi-definite square root $\Rmat_{nn}(f)^{1/2}$. If $\Rmat_{nn}(f)$ is further invertible\footnote{For the case where $\Rmat_{nn}(f)$ is non-invertible, see Appendix \ref{app:Rnn}.}, we can define:
\begin{equation}
  \label{eq:Qdef}
 \Qmat(f)=\Rmat_{nn}(f)^{-1/2}.
\end{equation}
By left-multiplication of (\ref{eq:audio_model}) by $\Qmat(f)$ we obtain
\begin{align}
  \Qmat(f)\mvect(f,t) &= \Qmat(f)\hvect(f,\thetavect)s(f,t) + \Qmat(f)\nvect(f,t), \\
  \label{eq:whitened}  
  \mvect'(f,t) &= \hvect'(f,\thetavect)s(f,t) + \nvect'(f,t),
\end{align}
where $\nvect'(f,t)$ follows the standard bivariate complex-normal $\mathcal{CN}_2(\zerovect,\Imat_2)$. Note that $\hvect'(f,\thetavect)$ can only be identified up to a multiplicative complex scalar constant because the same observations are obtained by dividing corresponding source signals by this constant. Hence, we can assume without loss of generality that $h'_1(f,\thetavect)=1$ and $h'_2(f,\thetavect)=r'(f,\thetavect)$, where $r'(f,\thetavect)$ is the relative transfer function (RTF) after whitening. It follows that, $m'_1(f,t)=s(f,t)+n'_1(f,t)$, $\sigma_{m'_1}^2(f,t)=\sigma^2_s(f,t)+1$ and $\sigma_{m'_2}^2(f,t)= |r'|^2 \sigma^2_s(f,t)+1$. Moreover, since $\Qmat(f)$ is invertible, the original RTF can be obtained from $r'(f,\thetavect)$ as the ratio of vector $\Qmat(f)^{-1}[1 , {r}'(f,\thetavect)]^\top$.

We can now use Theorem \ref{TH:MAIN} to obtain that $y'(f,t) = m'_2(f,t)/m'_1(f,t)$ follows the complex-t distribution:
\begin{equation}
\label{eq:biased_ratio}
\mathcal{CT}_1\hspace{-0.5mm}\left(\frac{\sigma^2_s(f,t)}{1+\sigma^2_s(f,t)}r'(f,\thetavect),\frac{\sigma_{m'_2}^2(f,t)+\sigma^2_s(f,t)}{(1+\sigma^2_s(f,t))^2},1\right)\hspace{-1mm}.
\end{equation}
Interestingly, it turns out that the distribution of a binaural ratio under white Gaussian noise is \textit{not} centered on the actual RTF $r'(f,\thetavect)$; but rather on a scaled version of it which depends on the instantaneous source variance. This suggests to use the following more natural feature that we refer to as \textit{rectified binaural ratio} (RBR):
\begin{equation}
\label{cor_bin_ratio}
y(f,t) = \frac{1+\sigma^2_s(f,t)}{\sigma^2_s(f,t)}\cdot\frac{m'_2(f,t)}{m'_1(f,t)}.\\
\end{equation}
This feature has the following distribution:
\begin{align}
\label{eq:pdfy}
P(y(f,t)) &= \mathcal{CT}_1\left(y(f,t);r'(f,\thetavect),\lambda^2(f,t),1\right), \\
\label{eq:lambda_case1}
\textrm{where}\hspace{2mm}\lambda^2(f,t) &= \frac{\sigma_{m'_2}^2(f,t)+\sigma^2_s(f,t)}{\sigma^4_s(f,t)},
\end{align}
which is centered on the RTF $r'(f,\thetavect)$. The spread parameter $\lambda^2(f,t)$ is also important because it models the uncertainty or ``reliability'' associated to each RBR feature: the larger is $\lambda^2(f,t)$, the less reliable is $y(f,t)$. Since the noise variance is fixed to 1, we see in (\ref{eq:lambda_case1}) that $\lambda^2(f,t)$ tends to 0 when the SNR at $(f,t$) tends to infinity, while $\lambda^2(f,t)$ tends to infinity when the SNR approaches 0, which matches intuition.

\vspace{-4mm}
\section{Parameter Estimation}
\vspace{-4mm}
\label{sec:param}
\subsection{Spread parameter}
\vspace{-2mm}
We consider the general case of time-varying source variances $\sigma^2_s(f,t)$. This is more challenging than a stationary model but also more realistic since typical audio signals such as speech or music are often sparse and impulsive in the time-frequency plane. In this case, the calculation of RBR features (\ref{cor_bin_ratio}) and of their spread parameter (\ref{eq:lambda_case1}) requires the knowledge of instantaneous source and microphone variances at each $(f,t)$.
A number of ways can be envisioned to estimate them. In this paper, we use the perhaps most straightforward approach: the instantaneous microphone variances $\sigma_{m'_1}^2(f,t)$ and $\sigma_{m'_2}^2(f,t)$ are approximated by their observed magnitudes $|m'_1(f,t)|^2$ and $|m'_2(f,t)|^2$. More accurate estimates could be obtained using, \textit{e.g.}, a sliding averaging window in the time-frequency plane as in \cite{vincent2009underdetermined}. However, this simple scheme showed good performance in practice. It leads to the following straightforward estimate for $\sigma^2_s(f,t)$:
\begin{equation}
\label{eq:missing}
\widehat{\sigma}^2_s(f,t) =
\left\{
 \begin{array}{l}
  |m'_1(f,t)|^2-1 \hspace{2mm}\textrm{if}\hspace{2mm} |m'_1(f,t)|^2  > 1,  \\
  0\hspace{2mm}\textrm{otherwise},
 \end{array}
\right.
\end{equation}
from which we deduce $\widehat{\lambda}^2(f,t)$ using (\ref{eq:lambda_case1}). $\widehat{\sigma}^2_s(f,t)=0$ leads to $\widehat{\lambda}^2(f,t)=+\infty$, corresponding to a missing data at $(f,t)$.

\vspace{-4mm}
\subsection{Unconstrained RTF}
\vspace{-2mm}
\label{subsec:freeh}
Once the spread parameter is estimated, we are left with the estimation of $r'(f,\thetavect)$ which is the mean of the complex t-distribution (\ref{eq:lambda_case1}). The  equivalent characterization of the t-distribution as a Gaussian scale mixture leads naturally to an EM algorithm that converges under mild conditions to the maximum likelihood \cite{McLachlanPeel2000b}. 
Introducing an additional set of latent variables $\uvect=\{u(f,t), f=1:F, t=1:T\}$, we can write (\ref{eq:pdfy}) equivalently as:
\begin{align}
P(y(f,t)|u(f,t)) &= \mathcal{CN}_1(y(f,t);r'(f,\thetavect),\frac{\lambda^2(f,t)}{u(f,t)}), \\
P(u(f,t))   &= \mathcal{G}(1,1),
% attention j'ai change les nu ici et avant pour etre coherent avec la litterature
\end{align}
where $\mathcal{G}$ denotes the Gamma distribution. At each iteration $(q)$, the M-step  updates $r'(f,\thetavect)$ as a weighted sum of the $y(f,t)$'s while the E-step consists of updating the weights 
defined as $\omega^{(q)}_{ft}=\frac{1}{2}\widehat{\lambda}^{-2}(f,t)\cdot\mathbb{E}[u(f,t)|y(f,t);r'^{(q)}(f,\thetavect)]$:
\begin{align}
 &\textbf{M-step:} \; r'^{(q+1)}(f,\thetavect) = (\textstyle\sum_{t=1}^T \omega^{(q)}_{ft} y(f,t))/(\textstyle\sum_{t=1}^T\omega^{(q)}_{ft}), \nonumber \\
 &\textbf{E-step:} \; \omega^{(q+1)}_{ft} = \left(\widehat{\lambda}^2(f,t) + |y(f,t)-r'^{(q+1)}(f,\thetavect)|^2\right)^{-1}. \nonumber
\end{align}
The initial weights $\omega^{(0)}_{ft}$ can be set to 1, although our experiments showed that random initializations usually converged to the same solution. Convergence is assumed reached when $r'(f,\thetavect)$ varies by less than $0.1\%$ at a given iteration. In practice, the algorithm converged in less than 100 iterations in nearly all of our experiments. Once an estimate $\widehat{r}'(f,\thetavect)$ is obtained, the non-whitened RTF $\widehat{r}(f,\thetavect)$ is calculated as the ratio of vector $\Qmat(f)^{-1}[1 , \widehat{r}'(f,\thetavect)]^\top$.

\vspace{-4mm}
\subsection{Acoustic space prior on the RTF}
\vspace{-2mm}
\label{subsec:RTF_prior}
In practice, when a sound source emits in a real room, the RTF can only take a restricted set of values belonging to the so-called \textit{acoustic space manifold} of the system \cite{deleforge2015acoustic}. Hence, a common approach is to search for the optimal $r'$ among a finite set of $K$ possibilities corresponding to different locations of the source, namely $r' \in \mathcal{R}' = \{r'_1, \ldots, r'_K\}$ where $r'_k(f) = r'(f, \thetavect_k)$. From a Bayesian perspective, this corresponds to a mixture-of-Dirac prior on $r'(f,\thetavect)$. Considering the observed features $y$, we then look for the $r'_{\widehat{k}}$ that maximizes the log-likelihood of $y$ as induced by (\ref{eq:pdfy}). Taking the logarithm of (\ref{eq:complex-t}),  this amounts to minimize:
\begin{equation}
 \widehat{k} = \operatorname*{argmin}_{k=1:K} \textstyle\sum_{t=1}^{T}\sum_{f=1}^{F}\log\bigl(\widehat{\lambda}^2(f,t) + |y(f,t) - r'_k(f)|^2\bigr). \nonumber
\end{equation}
We recover the robustness property that a data point with high spread has less impact on the estimation of $r'$.

\vspace{-4mm}
\section{Experimental results}
\vspace{-4mm}
\subsection{RTF estimation}
\begin{figure}[!t]
\centering
    \includegraphics[width = 0.95\linewidth,clip=,keepaspectratio]{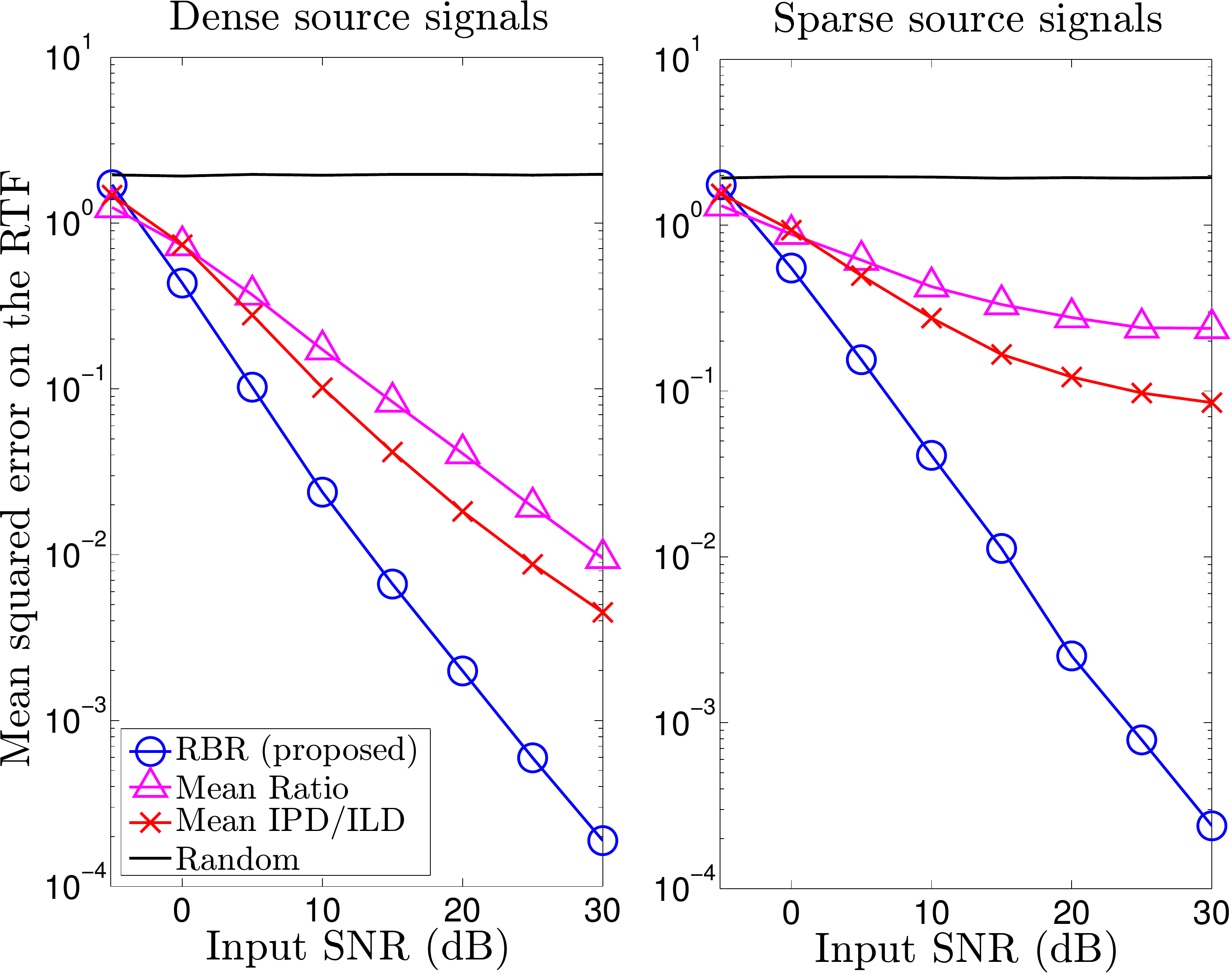}\vspace{-4mm}
    \caption{\small\label{fig:simu_results} Mean squared error of different RTF estimation methods for various SNRs.\vspace{-4mm}}
\end{figure}

\vspace{-2mm}
\label{subsec:xp_simu}
We first evaluate the RTF estimation method described in Section \ref{subsec:freeh} through extensive simulations. 160,000 binaural test signals are generated according to model (\ref{eq:audio_model}), (\ref{eq:stat_model_s}) and (\ref{eq:stat_model_n}), under a wide range of noise and source statistics. Each generated complex signal corresponds to $T=20$ time samples in a given frequency. The variances of source signals are time-varying and uniformly drawn at random. Sparse source signals are simulated by setting their variance to $0$ with a $50\%$ probability at each sample. For each test signal, the noise variances and correlation are uniformly drawn at random, and the RTF $r$ is drawn from a standard complex-normal distribution. The proposed method is compared to two baseline methods. The first one (Mean ratio) takes the mean of the complex microphone ratios $m_2(f,t)/m_1(f,t)$ over the $T$ samples of each signal. The second one (Mean ILD/IPD) calculates the mean ILD and IPD as follows:
\begin{equation}
\left\{
 \begin{array}{l}
 \overline{\textrm{ILD}}=\textstyle\frac{1}{T}\textstyle\sum_{t=1}^T\log\left(\frac{m_2(f,t)}{m_1(f,t)}\right), \\
 \overline{\textrm{IPD}}=\textstyle\frac{1}{T}\textstyle\sum_{t=1}^T\frac{m_2(f,t)/|m_2(f,t)|}{m_1(f,t)/|m_1(f,t)|}.
 \end{array}
\right.
\end{equation}
The RTF is then estimated as $\exp(\overline{\textrm{ILD}})\cdot\overline{\textrm{IPD}}$. This latter type of binaural cue aggregation is common to many methods, including \cite{Aarabi02,woodruff2012binaural,deleforge2015acoustic}. For fairness of comparison, the samples identified as \textit{missing} by our method according to (\ref{eq:missing}) are ignored by all 3 methods. Mean squared errors for various signal-to-noise ratios (SNR) and for both dense (left) and sparse (right) source signals are showed in Fig.~\ref{fig:simu_results}. As an indicator of the error upper-bound, the results of a method generating random RTF estimates (Random) are also shown.
Except at low SNRs ($\le$-5dB) where all 3 methods yield estimates close to randomness, the proposed method outperforms both the others. In particular, for SNRs larger than 15 dB, the mean squared error is decreased by  several orders of magnitudes and the RBR features performed best in $92\%$ of the tests. Two facts may explain these results. First, as showed in (\ref{eq:biased_ratio}), the microphone ratio is a biased estimate of the RTF under white noise conditions. This bias is further amplified for arbitrary noise statistics. Second, the baseline methods, as many existing methods in the literature, aggregate binaural cues with binary weights: each sample is classified as either missing or not. In contrast, the explicit spread parameter (\ref{eq:lambda_case1}) available for rectified binaural ratios enables to weight observations in a statistically sound way.

\vspace{-4mm}
\subsection{Time difference of arrival estimation}
\vspace{-2mm}
\begin{figure}[!t]
\centering
    \includegraphics[width = 0.9\linewidth,clip=,keepaspectratio]{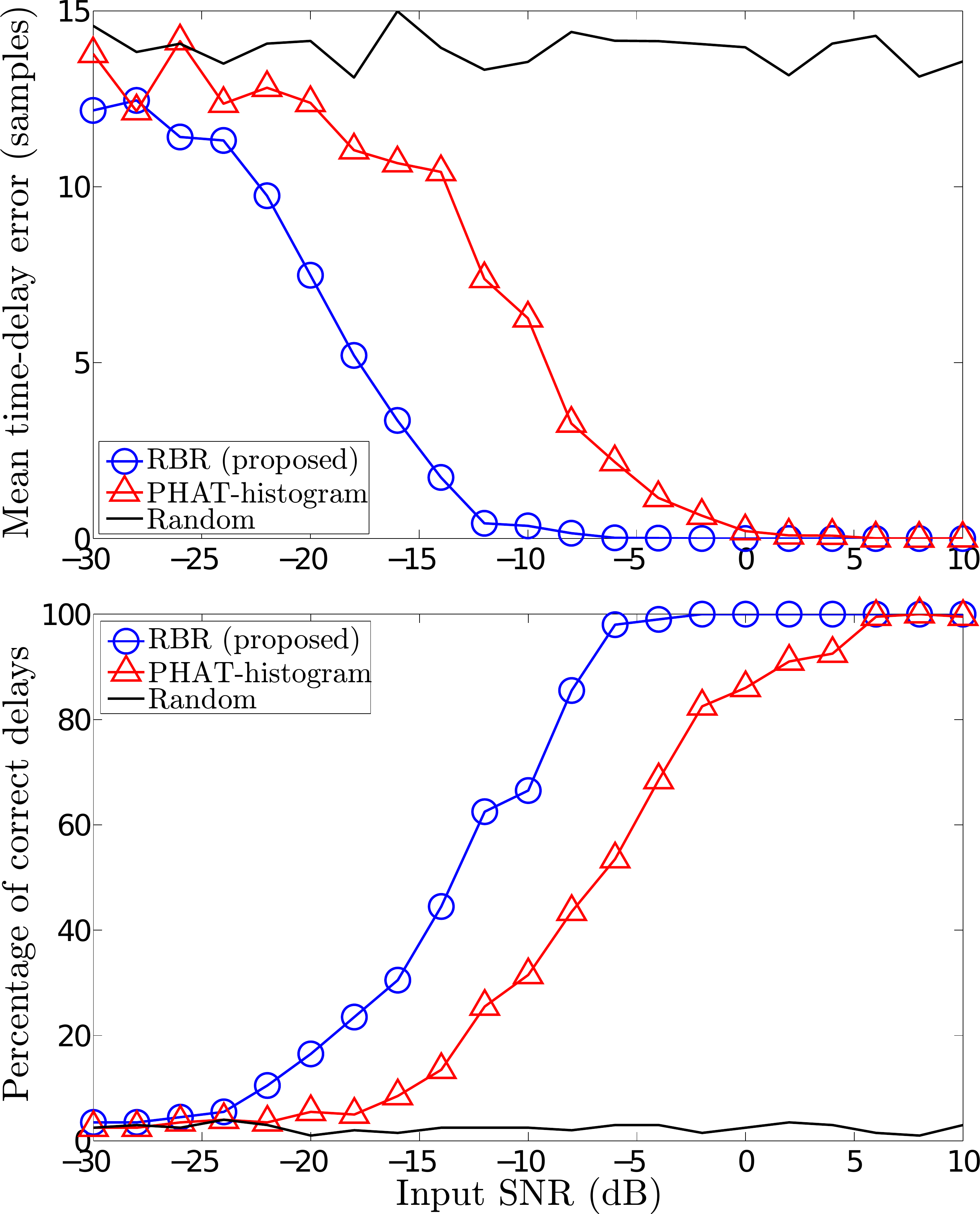}\vspace{-4mm}
    \caption{\small\label{fig:delay_results} Comparing time-delay estimation results of RBR and PHAT using 1 second noisy speech signals (200 test signals per SNR value).\vspace{-4mm}}
\end{figure}
Under free-field conditions, \textit{i.e.}, direct single-path propagation from the sound source to the microphones, localizing the source is equivalent to estimating the time difference of arrival (TDOA) between microphones. Indeed, for far enough sources, we have the relation $\tau \approx d\cos(\theta) F_s/C$ where $\tau$ is the delay in samples, $d$ the inter-microphone distance, $\theta$ the source's azimuth angle, $F_s$ the frequency of sampling, and $C$ the speed of sound. In the frequency domain, the RTF then has the explicit expression $r(f,\tau) = \exp(-2\pi i\tau (f-1)/F)$ where $F$ is the number of positive frequencies and $f=1:F$ is the frequency index. Let $\mathcal{R}$ be the discrete set of RTFs corresponding to delays of $-\tau_{\textrm{max}}$ to $+\tau_{\textrm{max}}$ samples, and $\mathcal{R}'$ the corresponding set after whitening, \textit{i.e.}, containing ratios of $\Qmat(f)[1, r(f,\tau)]^\top$. Given a noisy binaural signal, the method of Section \ref{subsec:RTF_prior} can be applied to select the most likely RTF $r'$ in $\mathcal{R}'$ and deduce the corresponding TDOA. $4,000$ test signals are generated using random 1 second speech utterances from the TIMIT dataset \cite{TIMIT} sampled at $F_s=$16,000 Hz. A binaural signal with a random delay of $-20$ to $+20$ samples between microphones is generated, before applying the short-time Fourier transform ($64$ms windows with $50\%$ overlap). This yields $F=512$ positive frequencies and $T=32$ time samples. These signals are finally corrupted by random additive stationary noise of known statistics in the frequency domain using the same procedure as in Section \ref{subsec:xp_simu}. The proposed RBR-based approach is compared to the sound source localization method PHAT-histogram\footnote{We used the PHAT-histogram implementation of Michael Mandel, available at http://blog.mr-pc.org/2011/09/14/messl-code-online/.}\cite{Aarabi02}. Results are displayed in Fig.~\ref{fig:delay_results}. For SNRs higher than -6 dB, the proposed RBR method yields less than $0.4\%$ incorrect delays, versus $10.1\%$ for PHAT-histogram on the same signals. RBR's average computational time is $80\pm6$ms per second of signal on a common laptop, which is about 3 times faster that PHAT-histogram using our Matlab implementations.

\vspace{-4mm}
\section{Conclusion}
\vspace{-4mm}
We explicitly expressed the probability density function of the ratio of two microphone signals in the frequency domain in the presence of a Gaussian-process point source corrupted by stationary Gaussian noise. This statistical framework enabled us to model the uncertainty of binaural cues and was efficiently applied to robust RTF and TDOA estimation. Future work will include extensions to multiple sound source separation and localization following ideas in \cite{mandel2010model}, and to more than two microphones following ideas in \cite{deleforge2015towards}. The flexibility of the proposed framework may also allow the inclusion of a variety of priors on the RTFs such as Gaussian mixtures, as well as the handling of various types of noise and source statistics.

\appendix
\vspace{-4mm}
\section{Appendix}
\vspace{-4mm}
\subsection{Non-invertible noise covariance}
\vspace{-2mm}
\label{app:Rnn}
If the noise signals $n_1(f,t)$ and $n_2(f,t)$ in (\ref{eq:audio_model}) have a deterministic dependency, $\Rmat_{nn}(f)$ is rank-1 and non-invertible. This is an important special case which may occur in practice when, \textit{e.g.}, the noise is a point source. Since $\Rmat_{nn}(f)^{-1/2}$ is then not defined, we replace the \textit{whitening} matrix in (\ref{eq:Qdef}) by
$
 \Qmat(f)=\left[
 \begin{array}{cc}
   1/\sigma_{n_1}^{2}(f) & 0 \\
   1/\sigma_{n_2}^{2}(f) & -1/\sigma_{n_2}^{2}(f) \\
 \end{array}
\right],
$
where $\sigma_{n_1}^{2}(f)$ and $\sigma_{n_1}^{2}(f)$ denote the variances of $n_1(f,t)$ and $n_2(f,t)$. It then follows that $n'_2(f,t)=0$ and that $n'_1(f,t)$ follows the standard complex-normal distribution $\mathcal{CN}(0,1)$. All subsequent derivations in the paper remain unchanged, with the exception of (\ref{eq:lambda_case1}) which becomes $\lambda^2(f,t) = \sigma_{m'_2}^2/\sigma^4_s$.

\vspace{-4mm}
\subsection{Proof of Theorem \ref{TH:MAIN} }
\vspace{-2mm}
\label{app:theo1}
We first prove the result for $\rho=0$, \textit{i.e.}, when $m_1$ and $m_2$ are independent. Since $[m_1,m_2]^T$ is jointly circular symmetric complex Gaussian, it follows that $m_1$ and $m_2$ are also complex Gaussian with $m_1 \sim  
\mathcal{CN}_1(0, \sigma^2_{m_1})$,  $m_2 \sim  
\mathcal{CN}_1(0, \sigma^2_{m_2})$  \cite{Gallager2008}, and $S^2=2 |m_1|^2/\sigma^2_{m_1}$ follows a Chi-square distribution with 2 degrees of freedom \cite{fuhrmann1997complex}. These properties generalize their counterparts in the real case and can be easily checked by using
 the characterization of complex Gaussians as real Gaussians on the real and imaginary parts \cite{Gallager2008}. We can now use the property of circular symmetric Gaussians that states that if $Y$ is $\mathcal{CN}(0, \Sigma)$ then $Y$ and
 $Y e^{i\phi}$ have the same distribution for all $\phi$. We deduce from this property that
 $y=m_2/m_1$ and $z= m_2/|m_1|$ have the same distribution. Then, $\sigma_{m_1} z = m_2\sqrt{2/S^2}$ is distributed as a complex Gaussian over the square root of an independent 
   scaled Chi-square distribution, which is one of the characterization 
  of the complex t-distribution\cite[Section 5.12]{KotzNadarajah2004} . It follows that  $\sigma_{m_1} z \sim
  \mathcal{CT}_1(0,\sigma^2_{m_2}, 1)$. Therefore $y$ follows $\mathcal{CT}_1(0,\sigma^2_{m_2}/\sigma^2_{m_1}, 1)$ which corresponds to Theorem \ref{TH:MAIN} for $\rho=0$.
  For the general case, we multiply $\mvect$ by matrix $\Amat=\left[
 \begin{array}{cc}
   1 & 0 \\
    -\rho^* & \sigma_{m_1}/\sigma_{m_2} \\
 \end{array}
\right]$
so that $\widetilde{\mvect} =\Amat \mvect$ is complex Gaussian with covariance matrix $ \Amat \Sigmamat \Amat^H= \sigma_{m_1}^2  \left[
 \begin{array}{cc}
   1 & 0 \\
    0 & 1-|\rho|^2 \\
 \end{array}
\right].$
We deduce from the previous case that $\widetilde{y}=\widetilde{m}_2/\widetilde{m}_1$ follows $\mathcal{CT}_1(0,1 - |\rho|^2, 1)$. We finally obtain Theorem \ref{TH:MAIN}'s result by noting that 
$\widetilde{y}=(\sigma_{m_1}/\sigma_{m_2})y - \rho^*$.
\small
\vspace{-4mm}
\bibliographystyle{IEEEbib_abrfname}
\bibliography{IEEEabrv,refs}

\begin{thebibliography}{10}

\bibitem{gannot2001signal}
S.~Gannot, D.~Burshtein, and E.~Weinstein,
\newblock ``Signal enhancement using beamforming and nonstationarity with
  applications to speech,''
\newblock {\em {IEEE} Trans. Signal Process.}, vol. 49, no. 8, pp. 1614--1626,
  2001.

\bibitem{knapp1976generalized}
C.~Knapp and G.~C. Carter,
\newblock ``The generalized correlation method for estimation of time delay,''
\newblock {\em {IEEE} Trans. Acoust., Speech, Signal Process.}, vol. 24, no. 4,
  pp. 320--327, 1976.

\bibitem{Aarabi02}
P.~Aarabi,
\newblock ``Self-localizing dynamic microphone arrays,''
\newblock {\em {IEEE} Trans. Syst., Man, Cybern. {C}}, vol. 32, no. 4, pp.
  474--484, 2002.

\bibitem{mandel2010model}
M.~I. Mandel, R.~J. Weiss, and D.~P. Ellis,
\newblock ``Model-based expectation-maximization source separation and
  localization,''
\newblock {\em {IEEE} Trans. Acoust., Speech, Signal Process.}, vol. 18, no. 2,
  pp. 382--394, 2010.

\bibitem{woodruff2012binaural}
J.~Woodruff and D.~Wang,
\newblock ``Binaural localization of multiple sources in reverberant and noisy
  environments,''
\newblock {\em {IEEE} Trans. Acoust., Speech, Signal Process.}, vol. 20, no. 5,
  pp. 1503--1512, 2012.

\bibitem{zohny2014modelling}
Z.~Zohny and J.~Chambers,
\newblock ``Modelling interaural level and phase cues with student's
  t-distribution for robust clustering in {MESSL},''
\newblock in {\em International Conference on Digital Signal Processing (DSP)}.
  IEEE, 2014, pp. 59--62.

\bibitem{mandel2015enforcing}
M.~I. Mandel and N.~Roman,
\newblock ``Enforcing consistency in spectral masks using markov random
  fields,''
\newblock in {\em EUSIPCO}. IEEE, 2015, pp. 2028--2032.

\bibitem{deleforge2015acoustic}
A.~Deleforge, F.~Forbes, and R.~Horaud,
\newblock ``Acoustic space learning for sound-source separation and
  localization on binaural manifolds,''
\newblock {\em International journal of neural systems}, vol. 25, no. 01, pp.
  1440003, 2015.

\bibitem{fuhrmann1997complex}
D.~R. Fuhrmann,
\newblock ``Complex random variables and stochastic processes,''
\newblock {\em The Digital Signal Processing Handbook}, pp. 60--1, 1997.

\bibitem{vincent2009underdetermined}
E.~Vincent, S.~Arberet, and R.~Gribonval,
\newblock ``Underdetermined instantaneous audio source separation via local
  gaussian modeling,''
\newblock in {\em Independent Component Analysis and Signal Separation}, pp.
  775--782. Springer, 2009.

\bibitem{KotzNadarajah2004}
S.~Kotz and S.~Nadarajah,
\newblock {\em {Multivariate t Distributions and their Applications}},
\newblock Cambridge, 2004.

\bibitem{baxley2010complex}
R.~J. Baxley, B.~T. Walkenhorst, and G.~Acosta-Marum,
\newblock ``{Complex Gaussian ratio distribution with applications for error
  rate calculation in fading channels with imperfect CSI},''
\newblock in {\em Global Telecommunications Conference (GLOBECOM)}. IEEE, 2010,
  pp. 1--5.

\bibitem{McLachlanPeel2000b}
G.~McLachlan and D.~Peel,
\newblock ``{Robust mixture modelling using the {T} distribution},''
\newblock {\em Statistics and computing}, vol. 10, pp. 339--348, 2000.

\bibitem{TIMIT}
J.~S. Garofolo, L.~F. Lamel, W.~M. Fisher, J.~G. Fiscus, and D.~S. Pallett,
\newblock ``The {DARPA TIMIT} acoustic-phonetic continuous speech corpus
  {CD-ROM},''
\newblock Tech. {R}ep. NISTIR 4930, National Institute of Standards and
  Technology, Gaithersburg, MD, 1993.

\bibitem{deleforge2015towards}
A.~Deleforge, S.~Gannot, and W.~Kellermann,
\newblock ``Towards a generalization of relative transfer functions to more
  than one source,''
\newblock in {\em EUSIPCO}. IEEE, 2015, pp. 419--423.

\bibitem{Gallager2008}
R.~Gallager,
\newblock ``{Circularly-symmetric Gaussian random vectors},''
\newblock preprint, 2008.

\end{thebibliography}

\end{document}